\documentclass[aps,twocolumn,fleqn,endnote]{revtex4-2}

\usepackage{graphicx}
\usepackage{amsmath} 
\usepackage[charter]{mathdesign}
\usepackage[hidelinks]{hyperref}
\DeclareMathAlphabet{\pazocal}{OMS}{zplm}{m}{n}

\usepackage{isomath}

\def\i{\mathrm{i}}
\def\d{\mathrm{d}}
\def\e#1{\mathrm{e}^{{#1}}}

\def\del#1{\delta\hspace{-0.08em}#1}
\def\vector#1{\mathbfit{#1}}

\def\tint{\mbox{{\Large$\int$}}}
\def\tintt#1{\mbox{{\Large$\int\limits_{\scriptscriptstyle#1}$}}}
\def\tfrac#1#2{\mbox{{\Large$\frac{#1}{#2}$}}}
\def\tsum{\mbox{{$\sum$}}}

\def\mat#1{\boldsymbol{#1}}

\def\rl{r_{\rm l}}
\def\rr{r_{\rm r}}
\def\tl{t_{\rm l}}
\def\tr{t_{\rm r}}

\def\matrix#1{\left(\!\!\begin{array}{cccc}#1\end{array}\!\!\right)}
\def\trans{{\scriptscriptstyle\mathsf{T}}}

\def\qmat{\mat{Q}}
\def\qvec{\mat{q}}
\def\smat{\mat{S}}
\def\svec{\mat{s}}
\def\vvec{\mat{v}}
\def\wvec{\mat{w}}
\def\vvec{\mat{\phi}}
\def\dmat{\mat{D}}
\def\omat{\mat{1}}
\def\umat{\mat{U}}
\def\vvec{\mat{v}}
\def\psivec{\mat{\psi}}
\def\usmat{\umat_{\!s}}
\def\uqmat{\umat_{\!q}}
\def\sa{\sin\alpha}\def\ca{\cos\alpha}%
\def\sc{\sin\gamma}\def\cc{\cos\gamma}%
\def\saa{\sin^2\!\alpha}%
\def\ccc{\cos^2\!\gamma}%
\def\ken{k}

\begin{document}

\title{Time delays in anisotropic systems}
\author{Ulf Saalmann} 
\author{Jan M. Rost}
\affiliation{Max-Planck-Institut f{\"u}r Physik komplexer Systeme, 
N{\"o}thnitzer Str.\ 38, 01187 Dresden, Germany}
\date{\today}

\begin{abstract}\noindent
Scattering properties and time delays for general (non-symmetric) potentials in terms of the respective S-matrices are discussed paradigmatically in one dimension and in comparison to symmetric potentials.
Only for the latter the Wigner and Smith time delays coincide. 
Considering asymmetric potentials also reveals that only one version of S-matrices used in the literature (the one with reflection coefficients on the diagonal) generalizes to the asymmetric case.
Finally, we give a criterion how to identify a potential with intrinsic symmetry which behaves like an asymmetric one if it is merely offset from the scattering center.
\end{abstract}

\maketitle

\section{Introduction}
\noindent
Time delays related to scattering phases \cite{wi55,sm60} have been discussed for a long time in transport problems \cite{te16,roam+11}.
More recently, they have been addressed in acoustics \cite{pama+23}, electromagnetics \cite{pami21}, and from a fundamental perspective of quantum trajectories \cite{dupa+22}, 
and since about a decade in the context of photo-ionization by ultra-short laser pulses. 
Experimentally, photo-ionization time delays have been extracted from streaking the momenta of electrons released by a short XUV pulse with a moderate IR field \cite{scfi+10} or so called RABBIT measurements aiming at the same time-delay information of the released electron wave-packet by using IR sidebands of the XUV photo-ionizing pulse train \cite{klda+11,issq+17}.
The link of the photo-ionization time delay to the Wigner-Smith time delay from scattering theory, as well as the delays in general emerging from these setups have been a source of ongoing debate \cite{saro20,febe+22,elgr23}. 
This is not surprising since the setups are quite intricate and become even more cumbersome, if the long-range Coulomb interaction comes into play, which is the case for almost all experiments performed. 

Recent experimental advances have made it possible to measure time delays originating in photo-ionizing molecules \cite{hujo+16,fudo+20,rikl+21,hojo+21,heha+22,goji+22}, that is from anisotropic potentials. This success motivates to ask for the theoretical foundation of time delays and their formulation for general interactions, since almost always time delay and S-matrices are discussed in the context of single-centered, often spherically-symmetric potentials \cite{kh23}.

In the following we elucidate basic properties of time delays in the simplest setting which is general enough to be sensitive to the properties of anisotropic (and isotropic, parity-respecting) potentials.
Since characteristic features (such as the difference between proper and partial time delays) can only be uncovered in a system with at least two independent scattering channels, we do not investigate photo-ionization but scattering in one dimension from a generic short-range potential, a scenario which provides two scattering channels.
Such type of scattering is relevant in complex media, in wave-guides, or generally, for transport problems.

Additional motivation is provided by the fact, that symmetric potentials in 1D hide subtleties of scattering and related time delays in at least two aspects: (i) Two different versions of the S-matrix $\smat$ are pursued in the literature 
\cite{me98,bi94,noro96,nu00}, which have different eigenvalues. Yet, both fulfill the criteria for S-matrices, derived from overarching principles of flux conservation and time-reversal invariance (for a real potential), namely that $\smat$ is unitary and symmetric. However, only the version which is a symmetric matrix with respect to incoming and outgoing channels \cite{me98} remains symmetric in case of anisotropic potentials. (ii) Furthermore, without symmetric interaction, the two commonly used formulations of time delay, namely partial time delays and proper time delays do not agree, prompting the question what their respective meaning is.

Scattering in one dimension was mostly theoretically investigated \cite{bi94,noro96,nu00,bo08} long before time delays have become popular, however, to the best of our knowledge never with a discussion or even a focus on situations where the scattering potential is not symmetric.

\section{Scattering in 1D}
\noindent
For our context a potential $V(x)$ is short range if at large distances $x$ the solutions of the time-independent Schr\"odinger equation $[-\d^2/\d x^2 +2V(x)-2E]\psi(x)=0$ are free waves, $\psi(|x|\gg 1) \propto e^{\pm\i \ken x}$ with $\ken\,{=}\,\sqrt{2 E}>0$, see also App.\,\ref{sec:coul}. 
We will use atomic units $e\,{=}\,\hbar\, { = }\, m_{\rm e}\,{\equiv}\,1$ and consider for convenience a particle of mass $m_{\rm e}$, unless stated otherwise.

\subsection{The S-matrix and its parameterization}
\noindent
There are two channels in an 1D scattering scenario.
Most easily
\cite{Note1} 
they are described by reflection ($r$) and transmission ($t$) amplitudes for incoming waves from the left or the right side. Asymptotically those wave function read,
with $\ken\,{=}\,\sqrt{2 E}$ and $\{\lim_{x\to{-}\infty}\psi(x,E),\;\lim_{x\to{+}\infty}\psi(x,E)\}$,
\begin{subequations}\label{eq:reftra}\begin{align}
\psi_{\rm l}(x,E) & =\{\e{+\i \ken x}{+}\rl(E)\,\e{-\i \ken x},\;\tl(E)\e{+\i \ken x}\}
\\
\psi_{\rm r}(x,E) & =\{\tl(E)\,\e{-\i \ken x},\;\e{-\i \ken x}{+}\rr(E)\e{+\i \ken x}\}.
\end{align}\end{subequations}
By means of the four reflection and transmission amplitudes in Eq.\,\eqref{eq:reftra} 
one gets immediately the scattering matrix
\begin{equation}\label{eq:smat}
\smat(E) = \matrix{\rl(E) & t_{l}(E) \\ \tr(E) & \rr(E)}\,.
\end{equation}
This S-matrix connects the amplitudes for incoming ($a_{\rm l,r}$) and outgoing ($b_{\rm l,r}$) waves 
\cite{Note2}
\begin{equation}\label{eq:b2a}
 \matrix{b_{\rm l}\\ b_{\rm r}}=\smat\matrix{a_{\rm l}\\ a_{\rm r}}.
\end{equation}
For both channels (l, r) particle-flux conservation guarantees $|r(E)|^{2}+|t(E)|^{2}=1$.
Furthermore, for real potentials $V(x)$ the complex-conjugated channels $\psi_{\rm l,r}^{*}$ are valid solutions as well.
Comparing $[\psi_{\rm l}^{*}{-}\rl^{*}\psi_{\rm l}]/\tl^{*}$ with $\psi_{\rm r}$ implies
 $\tl=\tr\equiv t$ and $\rl/\tl={-}\rr^{*}/\tr^{*}$. 
Those four conditions render the scattering matrix unitary ($\smat^{\dagger}\smat\,{=}\,\omat$) and symmetric ($\smat^{\trans}{=}\,\smat$). 

Note that the diagonal elements of $\smat$ connect the in- and outgoing channels on the same side (l$\to$l, r$\to$r). 
There is another version in the literature \cite{bi94,noro96,nu00} that mixes the channels on the diagonal (l$\to$r, r$\to$l). If the scattering potential is symmetric, the channel-mixing version fulfills all symmetries required for an S-matrix, namely that $\smat$ is unitary and symmetric. However, if the potential does not have parity,
this channel-mixing version is no longer symmetric and therefore is not a proper S-matrix. Since the overwhelming majority of published work discusses only symmetric potentials, this shortcoming of the channel-mixing S-matrix has not been pointed out.

For the unitary and symmetric S-matrix, three real parameters are sufficient
to define reflection and transmission amplitudes in the general form
\begin{subequations}\label{eq:parameter}\begin{align}
t(E) & = \cos(\alpha(E))\,\e{\i\beta(E)}
\\
r_{\rm l,r}(E) & = \i\,\sin(\alpha(E))\,\e{\i[\beta(E)\pm\gamma(E)]}.
\end{align}\end{subequations}
The basis $\psi_{\rm l,r}$ of incoming waves from the left and the right is only one of infinitely many choices. The one which diagonalizes the S-matrix with a suitable unitary transformation $\umat_s$ stands out and has the physical meaning, that the ratio of incoming waves from the right and left is not changed for the outgoing waves upon scattering. Due to the properties of the S-matrix, its diagonal representation $\dmat[\e{\i \svec(E)}]=\usmat{\!\!}^{\dagger}\!(E)\,\smat(E)\,\usmat(E)$ contains the eigenvalues given by pure phases $\vector s(E) = (s_1\;s_2)^\trans$. They read in terms of reflection and transmission amplitudes or in their parameterization \eqref{eq:parameter}
\begin{subequations}\label{eq:sdiag}
\begin{align}
\e{\i\,s_{1,2}(E)} & = \tfrac{\rl+\rr}{2} \pm\sqrt{\Big[\tfrac{\rl-\rr}{2}\Big]^2+t^2}
\\
&=\e{\i[\beta+\chi_\pm]}
\\
\chi_\pm&\equiv\mbox{atan}(\pm\sqrt{1{-}\sin^2\!\alpha\cos^2\!\gamma},\sin\alpha\cos\gamma).
\end{align}
\end{subequations}
If the potential is symmetric, i.e., $V(-x) = V(x)$, then $\rl=\rr$
and therefore $\gamma(E)=0$. 

We close this section with a note on the channel-mixing S-matrix \cite{bi94,noro96,nu00}: Even if it is a valid S-matrix (i.\,e.\ for the case of a symmetric potential where it is symmetric) its eigenvalues differ from those of the proper S-matrix \eqref{eq:smat} while the time delays, to be discussed below, agree. 
Only the time delays must agree since they are observables based on Hermitian operators. 
The S-matrix itself is not Hermitian and therefore not an observable. 
It provides a description of scattering, whose parameterization can be done in different ways, as long as they are consistent with the fundamental properties of a collision process.

\subsection{Partial time delays}
\noindent
As it has become clear from its definition, the S-matrix connects the two input with the two output channels. 
This means that a wave-packet sent from one side splits upon scattering and leaves the interaction region towards both sides.
Any change in the incoming configuration will lead to a different partition of outgoing waves to the left and the right. 
Time delays, however, are meaningful for the channels which diagonalize the S-matrix and thereby keep the ratio of waves entering and leaving the scattering region from the left and the right the same. Consequently, partial time delays are defined as
\begin{equation}\label{eq:parttd}
\widetilde{\tau}_{\!j}(E)=\tfrac{\d}{\d E}s_{\!j}(E)
\end{equation}
with $s_{\!j}$ given in Eqs.\,\eqref{eq:sdiag}.

\subsection{Proper time delays}
\noindent
Another, potentially more intuitive way of defining a time delay is to consider the dwell time $\tau$, i.\,e.\ the reduced or increased time it takes a wave-packet to traverse the region of interaction $|x|\le X$ compared to the traversal time $\tau_0$ of a free wave-packet,
\begin{subequations}\label{eq:dwell}\begin{align}
\tau & =\tint\!\d t \tintt{|x|\le X}\!\!\!\d x\,\rho(x,t) - \tau_{0}
\\
&= \tint\!\d t\,t\;\big[J({+}X,t)-J({-}X,t)\big] - \tau_{0}\,,
\end{align}\end{subequations}
where $\tau_{0}$ are the corresponding integrals for the free wave-packet involving $\rho_0$ and $J_0$, respectively.
The expression (\ref{eq:dwell}b) follows from (\ref{eq:dwell}a) by means of the continuity equation
and contains the time-dependent current $J$ (or $J_0$) of the wave-packet at the left and right boundary of the scattering region.
Note that in 1D the current and the current density are the same.
If $X$ is chosen sufficiently large for the asymptotic description \eqref{eq:reftra} to be valid,
it is obvious that $\tau$, defined spectrally by $\psivec(E)$, can be calculated by means of the reflection and transmission coefficients or the S-matrix for any given wave-packet. 
Without details, which can be found elsewhere \cite{jawa88,nu00}, the result is
\begin{equation}\label{eq:dwellpsi}
\tau = \tint\!\d E\, \psivec^{*}(E)\qmat(E)\psivec(E),
\end{equation}
with Smith's life-time matrix \cite{sm60}
\begin{equation}
\qmat(E) =-\i\,\smat^{\dagger}(E)\tfrac{\d}{\d E}\smat(E),
\end{equation}
where the time delay $\tau$ appears now as an expectation value of $\qmat(E)$ with the state $\psivec(E)$.
Indeed, $\qmat$ is a Hermitian matrix since $\smat$ is unitary.
Diagonalizing the life-time matrix $\qmat$
\begin{equation}\label{eq:qdiag}
\dmat[\qvec(E)]=\uqmat{\!\!}^{\dagger}\!(E)\,\qmat(E)\,\uqmat(E),
\end{equation}
yieds real eigenvalues which are called proper time delays
\begin{equation}\label{eq:proptd}
\overline{\tau}_{\!j}(E)=q_{\!j}(E).
\end{equation}
Since we consider a 2$\times$2 problem, they correspond to the minimal and maximal dwell time $\tau$ in Eq.\,\eqref{eq:dwellpsi} with the respective eigenvectors corresponding to the combination of incoming and outgoing waves that minimize or maximize the time delay. 

We have already introduced three different bases for the scattering channels, waves coming in from the left and the right to define the S-matrix in the first place, and the two bases which diagonalize the S-matrix giving partial time delays and the one which diagonalizes Smith's life-time matrix whose eigenvalues are the proper time delays. 
A fourth basis is often introduced if one wants to use a basis of real functions, namely linear combinations $\{\cos(\ken x),\sin(\ken x)\}$ of left- and right-traveling waves which have even and odd parity, respectively.
For completeness, we also give the S-matrix in the parity basis
\begin{equation}\label{eq:smatpar}
\smat_{\rm p}(E) = \e{\i\beta}\sin\!\alpha\matrix{ \i\,\cos\gamma{+}\cot\!\alpha & \sin\gamma \\ \sin\gamma & \i\,\cos\gamma{-}\cot\!\alpha}
\end{equation}
with details of the derivation given in App.\,\ref{sec:3d}.

Note that the S-matrix is always defined with respect to the center of incoming and outgoing waves. A potential 
with $V(-x) = V(x)$ we have called symmetric. There can be, however, the case, that the potential is symmetric
about about a point $x_{\rm cen}\ne 0$, which we call intrinsically symmetric. Scattering from such a potential will have formally a full S-matrix in any generic basis including the parity basis \eqref{eq:smatpar}. This prompts the question, if and how one could tell from experimental time delays, if the potential has intrinsic symmetry or not.
We will come back to this question later.

\section{Symmetry and time delays}
\subsection{Symmetric potentials and the Wigner-Smith time delay}\label{sec:sym}
\noindent
For a symmetric potential, $V(-x)=V(x)$, the S-matrix is diagonal
in the parity basis \eqref{eq:smatpar} since $\gamma = 0$. The eigenphases reduce to 
\begin{equation}\label{eq:svecsym}
\svec_{\rm sym}(E) = \matrix{\alpha(E)+\beta(E) \\ \pi-\alpha(E)+\beta(E)}.
\end{equation}
Also the life-time matrix becomes diagonal in this basis with the eigenvalues
\begin{equation}\label{eq:qvecsym}
\qvec_{\rm sym}(E) = \matrix{\alpha'(E)+\beta'(E) \\ -\alpha'(E)+\beta'(E)},
\end{equation}
where there prime denotes derivation with respect to $E$.
Hence, partial and proper time delays ($j{=}1,2$) agree for symmetric potentials
\begin{equation}\label{eq:ttsym}
\widetilde{\tau}_{\!j}(E)=\overline{\tau}_{\!j}(E).
\end{equation}
The equivalence holds similarly for spherical potentials in 3D, with the eigen-basis given by spherical harmonics.
It is this equivalence which has led to the notion of ``Wigner-Smith time delays''.

\subsection{Asymmetric potentials}
\noindent
For arbitrary potentials the equivalence \eqref{eq:ttsym} does not hold any longer.
With the eigenvalues from Eqs.\,\eqref{eq:sdiag} we obtain
\begin{subequations}\label{eq:levasy}\begin{equation}\label{eq:slevasy}
\widetilde{\tau}_{1,2}=\tfrac{\d}{\d E}s_{1,2}=\beta'\mp\frac{\cos\!\alpha\,\cos\gamma\,\alpha'-\sin\!\alpha\,\sin\gamma\,\gamma'}{\sqrt{1{-}\cos^{2}\!\gamma\,\sin^{2}\!\alpha}}
\end{equation}
and for the proper time delays from diagonalizing $\qmat$
\begin{equation}\label{eq:qlevasy}
\overline{\tau}_{1,2}=
q_{1,2}=\beta'\pm\sqrt{\alpha'{}^{2}+\sin^{2}\!\!\alpha\,\gamma'{}^{2}},
\end{equation}
\end{subequations}
where again we have dropped the energy dependence and indicate energy derivatives with a prime.
Obviously, these expression simplify to Eqs.\,\eqref{eq:svecsym} and \eqref{eq:qvecsym} for symmetric potentials ($\gamma{=}0$, $\gamma'{=}0$).

Although proper and partial time delays differ, the time delay averaged over the channels is the same for both
\begin{equation}\label{eq:ttsum}
\tfrac{1}{2}\tsum_{j=1,2}\widetilde{\tau}_{\!j}(E)
=\tfrac{1}{2}\tsum_{j=1,2}\overline{\tau}_{\!j}(E)
=\tfrac{\d}{\d E}\beta(E)\,.
\end{equation}
This result generalizes to higher dimensions (2D, 3D) where the sum runs formally over infinitely many channels, cf.\,App.\,\ref{sec:sum}. 

Since the eigenchannels of $\smat$ differ from those of $\qmat$, the natural question arises: What is the dwell time $\tau_{\!j}$ of a (shape-conserving) eigenchannel $j$ of $\smat$? 
Indeed the expectation value of $\qmat$ in an eigenfunction $\psivec_{\!j}$ of the partial time delay produces this partial time delay,
\begin{align}
\tau_{\!j}\equiv\psivec_{\!j}^{*}\qmat\psivec_{\!j}
=-\i\psivec_{\!j}^{*}\smat^{\dagger}\smat'\psivec_{\!j}=\widetilde\tau_{\!j}
\end{align}
as expected and shown in App.\,\ref{sec:dwellseig}.
\begin{figure}[b]
\includegraphics[width=\columnwidth]{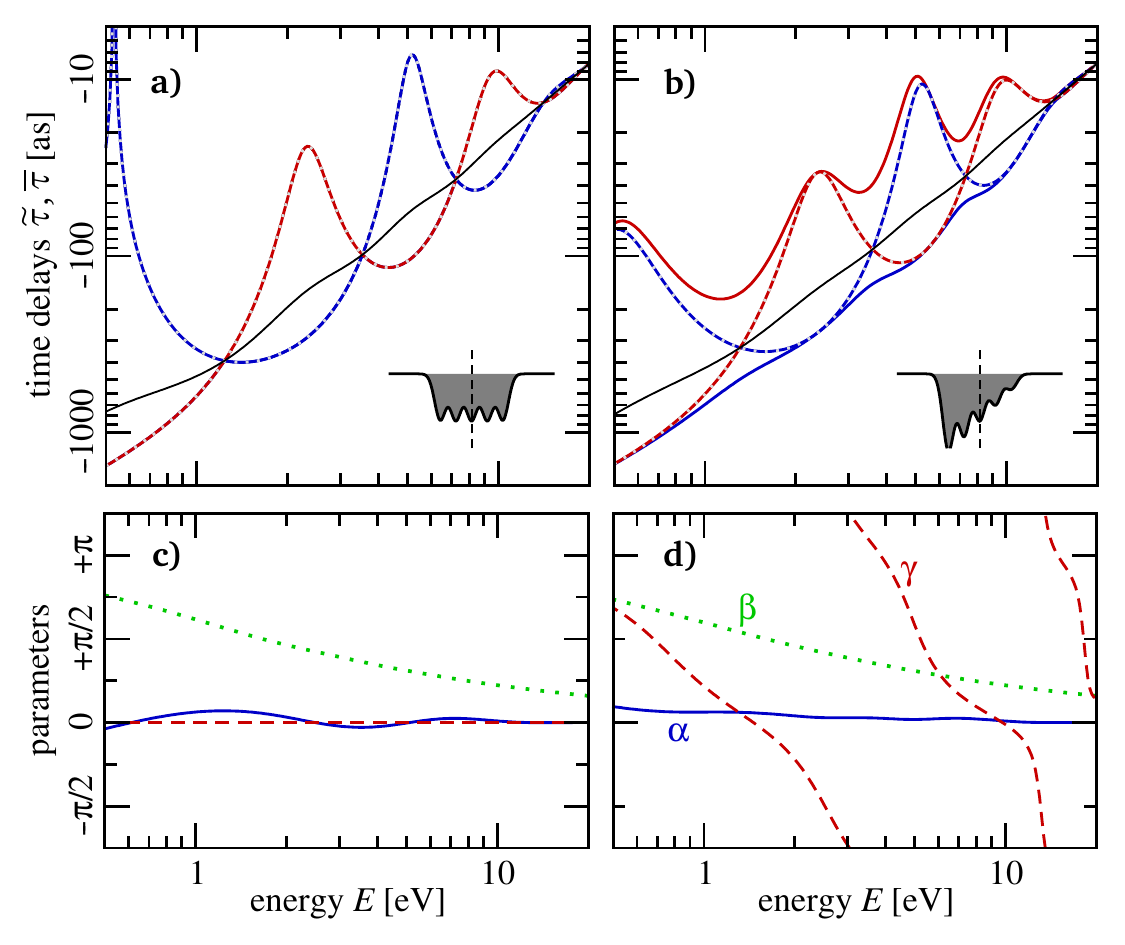}
\caption{Upper raw: Time delays for a symmetric ($V_{1}$, a) and an asymmetric ($V_{2}$, b) potential as a function of energy $E$, see Eq.\,\eqref{eq:pot12} and insets. Partial time delays $\widetilde{\tau}_{\!j}$ (dashed lines) according to Eq.\,\eqref{eq:parttd} are compared to proper time delays $\overline{\tau}_{\!j}$ (solid lines) given in Eq.\,\eqref{eq:proptd}. The averages of both $[\widetilde{\tau}_{1}{+}\widetilde{\tau}_{2}]/2$ and $[\overline{\tau}_{1}{+}\overline{\tau}_{2}]/2$ are identical (thin gray lines). Blue and red refer to the first and second channel, respectively.
\\
Bottom raw: The corresponding parameters $\alpha$ (solid-blue line), $\beta$ (green-dotted), $\gamma$ (red-dashed) as a function of energy $E$, cf.\,Eqs.\,\eqref{eq:parameter}.}
\label{fig:pot0102}
\end{figure}%

\subsection{Generic examples}
\noindent
We illustrate the time-delay behavior for generic symmetric and asymmetric potentials with two examples in Fig.\,\ref{fig:pot0102}. The potentials are defined by
\begin{equation}\label{eq:pot12}
V_{1,2}(x)=-V_{0}\sum_{j=-2}^{+2}f_{\!j}^{[1,2]}\,\e{-[x/d-2j]^{2}}
\end{equation}
with $V_{0}=2$\,eV and $d=1$\,\AA, to be specific.
The prefactors are $f_{\!j}{}^{[1]}\,{=}\,1$ and $f_{\!j}{}^{[2]}\,{=}\,1{+}j/3$ for the symmetric and asymmetric potential, respectively.

One can see in Fig.\,\ref{fig:pot0102}a that for the symmetric potential partial and proper time delays are identical for all energies $E$, as stated in Sect.\,\ref{sec:sym} above.

However, time delays differ for the asymmetric potential (Fig.\,\ref{fig:pot0102}b), where the proper time delays form an envelope for the partial time delays. 
This is to be expected as the eigenvalues of $\qmat$ are the minimal and maximal dwell times.
No scattering states (and therefore not even a shape-conserving one) can fall below or exceed those values.
Nevertheless, both time delays can agree at certain energies $E_{=}$, i.\,e., $\widetilde{\tau}_{\!j}(E_{=})=\overline{\tau}_{\!j}(E_{=})$, if
\begin{equation}
\alpha'=-\sin\!\alpha\,\cos\!\alpha\,\cot\!\gamma\,\gamma',
\end{equation}
as can be easily derived from Eqs.\,\eqref{eq:levasy}.
Similarly one can find those energies $E_{\times}$ where the partial time delays cross, i.\,e.\
$\widetilde{\tau}_{\!1}(E_{\times})=\widetilde{\tau}_{\!2}(E_{\times})$ which requires
\begin{equation}
\alpha'=\tan\!\alpha\,\tan\!\gamma\,\gamma'.
\end{equation}
For completeness we also present the energy dependence of the S-matrix parameters ($\alpha,\beta,\gamma$) in Figs.\,\ref{fig:pot0102}c,d, which confirm that $\gamma(E) = 0$ for a symmetric potential, but finite for an asymmetric one.

\section{The dependence of time delays on spatial properties}
\noindent
Time delay is not immune to shifting the potential in a coordinate system,
which may be surprising given its relative character, that is a delay relative to free motion at a given energy. 
Yet, scattering and subsequently time delay define a coordinate system, in particular a scattering center through incoming and outgoing waves and the S-matrix.
The location of the potential relative to the scattering center will have an influence on the time delay.
A very loose analogy is the echo of an object placed at some distance in front of a reflecting wall, which plays the role of the scattering center:
The echo one receives will depend on the object as well as on its distance to the wall. Similarly, for angular differential cross sections parameterized with partial waves, the amplitudes of the partial waves depend on the location of the origin of the coordinate system relative to the target.
We will elucidate the dependence of the time delays on spatial properties of the scattering scenario below with examples.

\subsection{Position of the potential with respect\\ to the scattering state}
\noindent
The results so far render time delay a useful observable, if carefully assessed in a specific physical situation.
What makes time delay, however, quite cumbersome is the fact that 
it depends also on the location of the potential with respect to the incoming wave-packet \cite{pama+23}. While it is natural (albeit not necessary) to place an intrinsically symmetric potential at the origin rendering it symmetric, no obvious choice exists a priori for an asymmetric potential. 
In a realistic situation time delays are extracted from asymptotic electron wave-packets. 
Those wave-packets have a clear origin, but it is the location of the potential (e.\,g.\ the molecule) with respect to that origin which matters for the time delays.
This location is not uniquely ``defined'' and could be difficult to determine.

\begin{figure}[b]
\includegraphics[width=\columnwidth]{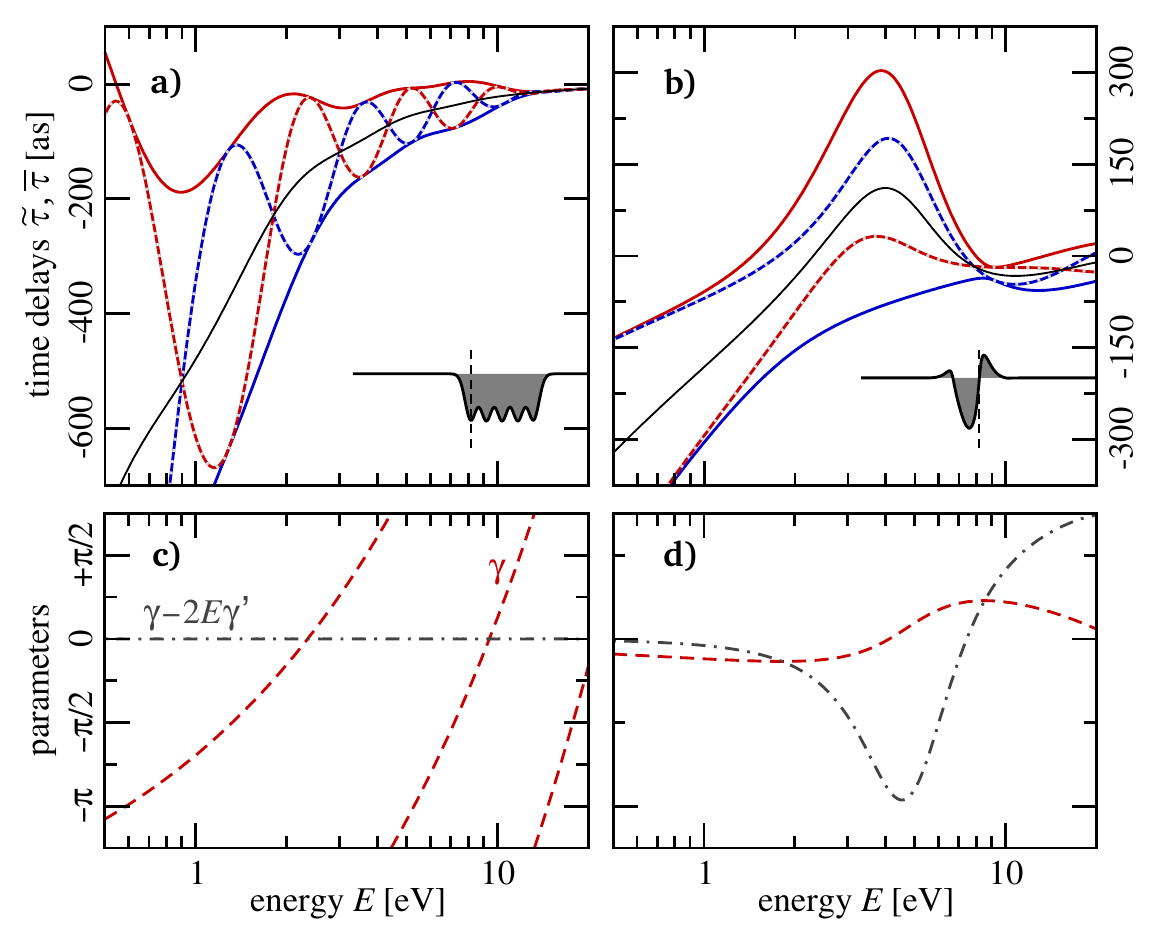}
\caption{Upper row: Time delays for a displaced potential ($V_{3}$, a) and a potential with a resonance ($V_{4}$, b) as a function of energy $E$, see Eqs.\,\eqref{eq:pot3}, \eqref{eq:pot4} and insets. See caption of Fig.\,\ref{fig:pot0102} for line styles.
\\
Bottom row: Corresponding parameters $\gamma$ (red-dashed) and $\gamma{-}2E\gamma'$ (gray-dot-dashed) as a function $E$.
Note that for the symmetric potential (panel c) $\gamma\sim \sqrt{E}$ and thus $\gamma{-}2E\gamma'\sim\mbox{const}$. 
}
\label{fig:pot1133}
\end{figure}

One should note, however, that general time delays in the absence of particular spectral features have only become of interest with the advent of ultra-short laser pulses. 
Before, time delays were mostly discussed in relation to a resonance.
At the resonance energy, the time delays are drastically enhanced for all channels sensitive to the resonance, see Fig.\,\ref{fig:pot1133}b, where time delays are shown for the potential
\begin{equation}\label{eq:pot4}
V_{4}(x)=\e{-[x/d-1/2]^{2}}\mbox{atan}\big(2\sin(2x/d)\big).
\end{equation}
In such a situation the location of the potential plays a subdominant role.
This is probably the reason, that the difference between various time delay definitions and the dependence of those time delays on the location of the potential has seen little attention to date.

However, the consequence of shifting the potential can be significant and is particularly dramatic for a 
 symmetric potential. We shift the one defined in Eq.\,\eqref{eq:pot12} according to
\begin{equation}\label{eq:pot3}
V_{3}(x)=V_{1}(x-\del x)
\end{equation}
which turns it from a symmetric one into one with only intrinsic symmetry. The result is shown for a particular displacement of $\del{x} = 2d$ in Fig.\,\ref{fig:pot1133}a. Not only do proper and partial time delays no longer agree, also the partial time delays differ substantially from the ones where the location of the potential is chosen such that it becomes symmetric as in Fig.\,\ref{fig:pot0102}a. Yet, the intrinsic symmetry of the potential is
still reflected in a more subtle relation as one can see in Fig.\,\ref{fig:pot1133}c, namely that $\gamma(E)\,{-}\,2E\gamma'(E)= 0$ holds for all energies $E$.

\begin{figure}[t]
\includegraphics[width=\columnwidth]{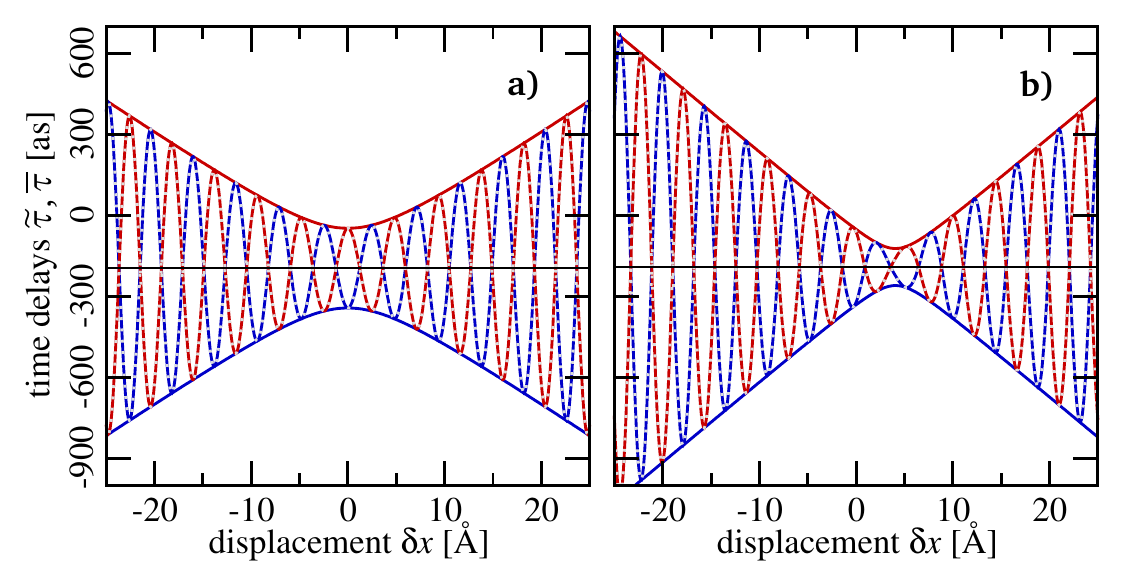}
\caption{Time delays for a symmetric ($V_{1}$, a) and an asymmetric ($V_{2}$, b) potential as a function a displacement $\del{x}$ with respect to the scattering center, see Eq.\,\eqref{eq:pot12} and insets. The energy is $E=2$\,eV in both cases. See caption of Fig.\,\ref{fig:pot0102} for line styles.}
\label{fig:pot0102-shift}
\end{figure}%
This relation, which is not fulfilled for a generically asymmetric potential, see Fig.\,\ref{fig:pot1133}d, can be understood from \eqref{eq:parameter} and the following consideration regarding the effect on the time delays when shifting the potential by $\del x$, see Fig.\,\ref{fig:pot0102-shift}.
 A little thought reveals that shifting the potential's position by $\del{x}$ will only change $\gamma$ according to 
\begin{subequations}\label{eq:gamdelx}\begin{align}
\gamma_{\!\del{x}}(E)&=\gamma_0(E)+2\ken\del{x},
\intertext{such that}
\gamma'_{\!\del{x}}(E)&=\gamma'_0(E)+2\del{x}/\ken,
\end{align}\end{subequations}
where $\gamma_0(E)$ refers to some reference position of the potential.

Partial and proper time delays can still be obtained from Eqs.\,\eqref{eq:levasy}
with $\gamma$ and $\gamma'$ replaced by $\gamma_{\!\del{x}}$ and $\gamma'_{\!\del{x}}$, respectively.
The difference $\Delta\overline{\tau}=\overline{\tau}_1-\overline{\tau}_2$ grows for large displacement $\del{x}$ linearly with a slope of $2/\ken$. Taking into account $\hbar$, the slope defines a velocity, which (multiplied
with $\del{x}$) represents the increased/reduced time a particle needs to reach/leave a displaced potential. 

The minimal ``gap'' $\Delta\overline{\tau}_{\rm min}=2\alpha'$ of the proper time delays occurs according to Eqs.\,(\ref{eq:levasy}b) and (\ref{eq:gamdelx}b) for $\gamma'_{\del{x}_{\rm min}}=0$, realized with $\del{x}_{\rm min}=-\ken\gamma'_0/2$. 
The resulting 
\begin{equation}\label{eq:gamdelxmin}
 \gamma_{\!\del{x}_{\rm min}}=\gamma_0-\ken^2\gamma'_0
\end{equation}
varies for generically asymmetric potentials with energy in the expression for the partial-time-delay difference
\begin{equation}
 \Delta\widetilde{\tau}_{\rm min} =2\frac{\cos\!\alpha\,\cos(\gamma_0{-}\ken^2\gamma'_0)\,\alpha'}{\sqrt{1{-}\sin^{2}\!\alpha\,\cos^{2}(\gamma_0{-}\ken^2\gamma'_0)}} 
\end{equation}
at the minimal gap of the proper time delays.
Figure~\ref{fig:pot0102-shift} illustrates this for a particalur energy. 

For potentials with intrinsic symmetry the reference in Eqs.\,\eqref{eq:gamdelx} can be chosen such that $\gamma_0\,{=}\,0$ (by making $x_{\rm cen}\,{=}\,0$), which entails $\gamma'_0\,{=}\,0$.
Once this potential is offset from the center, $\gamma$ becomes finite and energy-dependent.
Yet, as follows directly from \eqref{eq:gamdelx}, $\gamma_{\del{x}}{-}\,2E\gamma'_{\del{x}}\,{=}\,\mbox{const}$ for all energies $E$.
For a generically asymmetric potential (not possessing an $x_{\rm cen}$) this is not possible.
 
We may conclude that the interplay of partial and proper time delays reveal the symmetry of the potential, despite their sensitivity to its location: If the minimal gap of the proper time delays coincides at all energies with the maximal gap of the partial time delays, the underlying potential has intrinsic symmetry.

\subsection{Reflection-less potentials}
\begin{figure}[h]
\includegraphics[width=\columnwidth]{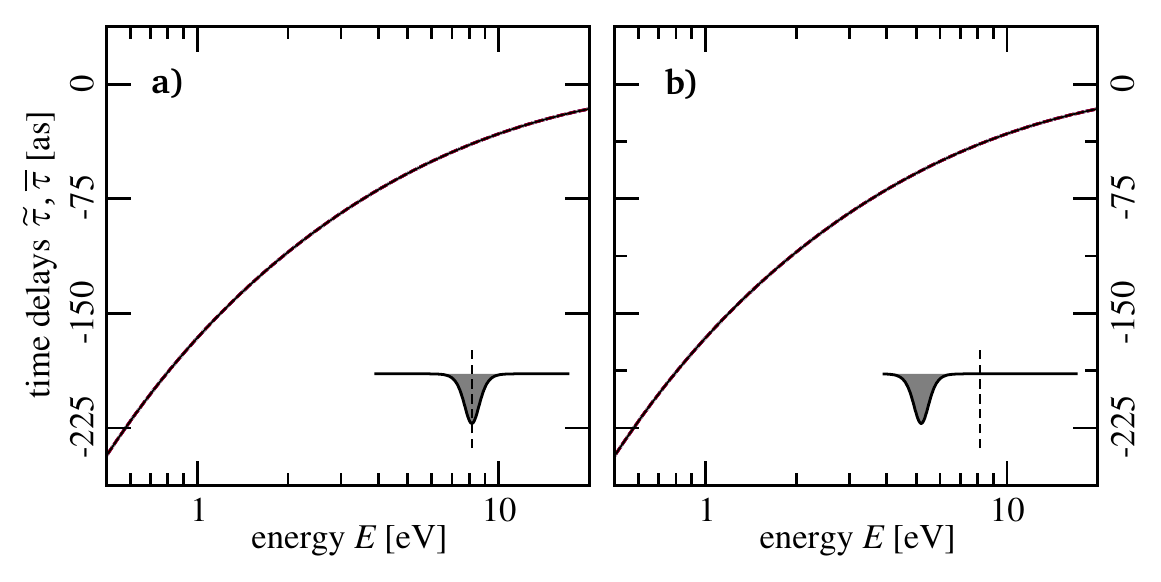}
\caption{Time delays for a reflection-less potential ($V_{5}$, a) and the same one displaced ($V_{6}$, b) , cf.\ Eq.\,\eqref{eq:pot5} and see insets, as a function of energy $E$. See caption of Fig.\,\ref{fig:pot0102} for line styles.}
\label{fig:pot0414}
\end{figure}

\noindent
Apparently, time delays are quite sensitive to the quantum-mechanical interference of transmitted and reflected waves. Hence, one would expect a radically different behavior, if reflection is suppressed. This can be double-checked by investigating the time delays of a reflection-less potential.
It is well known \cite{to97} that certain potentials show perfect transmission $|t(E)|^{2}=1$ for all energies $E$,
e.g.,
\begin{equation}\label{eq:pot5}
V_{5}(x)=\tfrac{1}{d^{2}\cosh^{2}(x/d)}\,.
\end{equation}
In Fig.\,\ref{fig:pot0414} we show partial and proper time delays for $V_{5}(x)$ and its shifted version $V_{6}(x)=V_{5}(x-\del{x})$ with $\del{x}=3$\,\AA.
As can be seen, all time delays agree even for the displaced potential. This reveals that subtle interference effects
due to the position of the potential as well as the difference of proper and partial time delays are of pure quantum 
nature and vanish in a (semi-)classical setting, as provided by the potential free of reflections which behaves as a classical system would do (full transmission and zero reflection).
Of course, in such a situation only a single channel is left and therefore even in full quantum mechanics, no interference can occur. 
This lets the proper and partial time delays collapse to a single time delay which is identical to the average one.
All of this follows directly from the parameters \eqref{eq:parameter} in this case. It is $\alpha\,{=}\,0$, thus $\alpha'\,{=}\,0$ and $\gamma$ being irrelevant.

\section{Conclusion}
\noindent
We have discussed scattering and ensuing time delays for generic potentials (without symmetry), paradigmatically in one dimension. For potentials symmetric to the origin, the standard case almost exclusively discussed also in 3D with centrally-symmetric potentials,
proper and partial time delays agree which has given the notion of  Wigner-Smith time delays.
This is not the case for generic potentials whose location relative to the collision origin have significant influence on the various time delays. With asymmetric potentials, one also notices a widespread use of a definition of the S-matrix which is only symmetric (as it should be from general considerations) for symmetric potentials and therefore not suitable to describe scattering for generic potentials.
Finally, we have provided a criterion which allows one to identify from the time delays an intrinsically symmetric potential (symmetric with respect to some position $x_{\rm cen}$) located at an arbitrary position. 

\section*{Acknowledgment}\vspace*{-3ex}
\noindent
We dedicate this analysis to Ravi Rau, who has always been an inspiration and role model to view a problem from different perspectives and to articulate it in a clear and transparent manner. 
We wish him many productive years to come.

\def\thesection{\Alph{section}}\def\thesubsection{\arabic{subsection}}\setcounter{section}{0}
\section{Appendix}
\subsection{Coulombic systems}\label{sec:coul}
\noindent
Long-range Coulomb potentials lead to infinite time delays if referenced with free motion \cite{sm60}.
(Note, that finite values, reported in experiments \cite{scfi+10,klda+11}, are the consequence of the measurement \cite{saro20}.)
Therefore, free motion must be replaced by motion in a pure Coulomb potential as reference \cite{saro20}. 
Then time delays are finite and measure the (short-range) deviation from a pure Coulomb interaction.
In 3D, many of the ideas and techniques presented here carry over to Coulomb systems with asymptotic wavefunctions $\exp(\pm[\i\ken r+\frac{1}{\ken}\ln(2\ken r)])$, i.\,e., amended by the logarithmic Coulomb phase. 
However, dynamics with Coulomb interaction differs substantially in 1D and 3D and disguises the general analogy between 1D and 3D scattering.

\subsection{Analogy to 3D scattering in a basis\\ of spherical harmonics}\label{sec:3d}
\noindent
The treatment described so far is specific to 1D, but has a clear relation to the 3D situation:
The continuous solid angle $\{\vartheta,\varphi\}$ in 3D gets replaced in 1D by two discrete directions $d\,{=}\,{-}1$ (left) and $d\,{=}\,{+}1$ (right), respectively. 
The relation between 1D and 3D becomes transparent with the commonly-used spherical-harmonics basis $Y_{\ell m}(\vartheta,\varphi)$
with $\ell\,{=}\,0\ldots\infty$ and $m\,{=}\,{-}\ell\ldots{+}\ell$ for the angular degrees of freedom.
In 1D one can use two ($m\,{=}\,0,1$) ``angular'' functions $y_{m}(d)=d^{m}\!/\!\sqrt{2}$,
which are orthonormal $\sum_{d=\pm 1}y_{m}(d)y_{m'}(d)=\delta_{mm'}$.
Whereas the description with a finite $\ell_{\rm max}$ is approximate but numerically accurate in 3D,
the description in 1D in terms of the $y_{\!m}$ is exact.

Instead of the traditional form \eqref{eq:reftra} in the Cartesian coordinate $x$, one can write the two continuum states ($j{=}1,2$) asymptotically in terms of the $y_{m}(d)$ and real radial functions $\phi_{mj}(r)$ in the discrete ``angle'' $d\,{=}\,\mbox{sgn}(x)$ and the radial distance $r\,{=}\,|x|$, respectively, as
\begin{subequations}\label{eq:3dform}\begin{align}
\psi_{j}(r,d,E) & = \sum_{m=0,1}y_{m}(d)\phi_{mj}(r,E) \tag{\ref{eq:3dform}}
\\
\left.{\phi^\infty_{01}(r,E) \atop \phi^\infty_{12}(r,E)}\right\} & = \cos(\ken r)\pm\cos\alpha \cos(\ken r{+}\beta)
\notag\\[-1ex] &\qquad\pm\cos\gamma\sin\alpha\sin(\ken r{+}\beta)
\\
\left.{\phi^\infty_{11}(r,E)\atop\phi^\infty_{02}(r,E)}\right\} & = -\sin\gamma\sin\alpha\cos(\ken r{+}\beta),
\end{align}\end{subequations}
where we have omitted the dependence of $\alpha$, $\beta$ and $\gamma$ on the energy $E{=}\ken^{2}\!/2$ and used the notation $\phi^\infty(r,E)\equiv\lim_{r\to\infty}\phi(r,E)$.
In general, in both functions ($j\,{=}\,1,2$) the two ``angular'' channels ($m\,{=}\,0,1$) couple, as can be seen in Eqs.\,\eqref{eq:3dform}.
For symmetric potentials, where $\gamma=0$ [cf.\,Eqs.\,\eqref{eq:parameter}], this is not the case leading to the simplification
\begin{subequations}\label{eq:3dformsym}\begin{align}
\left.{\phi^\infty_{01}(r,E) \atop \phi^\infty_{12}(r,E)}\right\} & = \cos(\ken r)\pm\cos(\ken r{\pm}\alpha{+}\beta)
\\
\left.{\phi^\infty_{11}(r,E)\atop\phi^\infty_{02}(r,E)}\right\} & = 0,
\end{align}\end{subequations}
where one can directly read off the eigenphases \eqref{eq:svecsym}.
Note, that radial wave-packets built from either symmetric ($m{=}0$) or anti-symmetric ($m{=}1$) states, given in Eq.\,(\ref{eq:3dformsym}a), will keep their symmetry throughout the scattering process.

The asymptotic expressions of the radial functions \eqref{eq:3dform} have the form
\begin{equation}
 \phi^\infty_{mj}(r,E) = a_{mj}\e{-\i \ken r}+b_{mj}\e{+\i \ken r},
\end{equation}
from which the S-matrix is directly obtained by means of Eq.\,\eqref{eq:b2a}.
Since we have used the parity basis $y_{m=0,1}$ we get $\smat$ given in Eq.\,\eqref{eq:smatpar}.

In 3D the analogous form of Eqs.\,\eqref{eq:3dform} is the most efficient way to calculate the continuum functions at energy $E$, since an equivalent form of Eqs.\,\eqref{eq:reftra} is not available.
The calculations can be done by means of the coupled-channel renormalized Numerov method \cite{jo78}.

\subsection{Sum of time delays}\label{sec:sum}
\noindent
It is shown that the sum of proper ($\overline{\tau}_{\!j}$) and partial ($\widetilde{\tau}_{\!j}$) time delays are equal \cite{te16}.
We will use repeatedly the unitarity of $\usmat$ as well as $\smat$ and the possibility to change the order of matrices under the trace.
Therewith
\begin{align}\label{eq:sum}
 \sum_{j}\overline{\tau}_{\!j}& =
\sum_j q_j=\mathrm{tr}(\uqmat{\!\!}^{\dagger}\!\qmat\uqmat)
=\mathrm{tr}(\qmat)
=-\i\,\mathrm{tr}\big(\smat^{\dagger}\smat'\big)
\notag\\[-2ex]
&=-\i\,\mathrm{tr}\Big(\usmat\dmat[\e{-\i\svec}]\usmat{\!\!}^{\dagger}
 \big(\usmat\dmat[\e{\i\svec}]\usmat{\!\!}^{\dagger}\big)'\Big)
\notag\\ &
=-\i\,\mathrm{tr}\big(\dmat[\e{-\i\svec}]\dmat'[\e{\i\svec}]\big)
-\i\,\mathrm{tr}\big(\usmat{\!\!}^{\dagger}\usmat'+(\usmat{\!\!}^{\dagger})'\usmat\big)
\notag\\ &=\mathrm{tr}\big(\dmat[\e{-\i\svec}]\dmat[\svec'\e{\i\svec}]\big)
-\i\,\mathrm{tr}\big((\usmat{\!\!}^{\dagger}\usmat)'\big)
\notag\\
&=\sum_{\!j}s'_{\!j} = \sum_{j}\widetilde{\tau}_{\!j}.
\end{align} 
As above, $\dmat[]$ denotes a diagonal matrix.

Note that we have nowhere used the fact that we treat a 1D system with 2$\times$2 matrices.
Thus Eq.\,\eqref{eq:sum} holds for any dimension.

\subsection{Dwell time for a scattering eigenchannel}\label{sec:dwellseig}
\noindent
We assume that $\vvec_{\!j}$ is an eigenvector of the scattering matrix $\smat\vvec_{\!j}=\vvec_{\!j}s_{\!j}$.
Therefrom follows
\begin{align}\label{eq:sprime}
s'_{\!j}
&=\vvec_{\!j}{}'\smat\vvec_{\!j}+\vvec_{\!j}\smat\vvec'_{\!j}+\vvec_{\!j}\smat'\vvec_{\!j}
\notag\\
& = s_{j}[\vvec_{\!j}{}'\vvec_{\!j}+\vvec_{\!j}\vvec'_{\!j}]+\vvec_{\!j}\smat'\vvec_{\!j}
=\vvec_{\!j}\smat'\vvec_{\!j}
\end{align}
since $\vvec_{\!j}\vvec_{\!j}\,{=}\,1$ and thus the term in brackets vanishes.
Therewith we can calculate the expectation value of $\qmat$ for the scattering eigenstate
\begin{align}
\tau_{\!j}&\equiv\vvec_{\!j}\qmat\vvec_{\!j}
=-\i\vvec_{\!j}\smat^{\dagger}\smat'\vvec_{\!j}
\notag\\
&=-\i\; s_{\!j}\vvec_{\!j}\smat'\vvec_{\!j}
=-\i\; s_{\!j}s'_{\!j}=\widetilde{\tau}_{\!j},
\end{align}
where in the 2nd line we have used Eq.\,\eqref{eq:sprime}.

\subsection{Matrices, eigenvalues and eigenvectors\\ in paramterized form}
\noindent
For completeness, the matrices $\smat$ and $\qmat$ and their eigenforms are given in terms of the parametrization \eqref{eq:parameter}
\begin{subequations}\label{eq:smatabc}\begin{align}
\smat&=\matrix{\i\e{\i[\beta+\gamma]}\sa& \e{\i\beta}\ca\\ \e{\i\beta}\ca & \i\e{\i[\beta-\gamma]}\sa}
\tag{\ref{eq:smatabc}}
\\
s_{1,2} 
&=\e{\i[\beta+\mbox{atan}(\pm\sqrt{1-\ccc\saa},\cc\sa)]}
\\
\vvec_{\!1,2} &=\frac{1}{N_s}\matrix{-\sa\sc\pm\sqrt{1-\ccc\saa}\\ \ca}
\end{align}\end{subequations}
with $N_s$ ensuring normalization.
Note that the eigenvectors of $\smat$ can be chosen real (which is generally not the case for unitary matrices). 
And further
\begin{subequations}\label{eq:qmatabc}\begin{align}
\qmat&=\matrix{+\saa\,\gamma'& \e{-\i\gamma}[\alpha'-\i\,\eta]\\ \e{+\i\gamma}[\alpha'+\i\,\eta] & -\saa\,\gamma'}
\tag{\ref{eq:qmatabc}}
\\
& \qquad\eta\equiv \ca\sa\,\gamma'\notag
\\
q_{1,2} &=\beta'-\sqrt{\alpha'{}^2+\saa\gamma'{}^2}
\\
\wvec_{\!1,2} &=\frac{1}{N_q}\matrix{\pm\sqrt{\alpha'{}^2+\saa\,\gamma'{}^2}+\saa\,\gamma'\\ \e{\i\gamma}[\alpha'+\i\,\ca\sa\,\gamma']}
\end{align}\end{subequations}
again with the normalization $N_q$ not explicitly given.
The unitary matrices referred to in the text are $\usmat\,{=}\,(\vvec_{\!1}\;\vvec_{\!2})$ and $\uqmat\,{=}\,(\wvec_{\!1}\;\wvec_{\!2})$.

\vfill

\def\articletitle#1{\emph{#1.}}

\end{document}